\documentclass[pra,aps,showpacs,preprint]{revtex4}
\usepackage{dcolumn}

\begin{document}

\topmargin 0.0in

\title{Extra dimensions present a new flatness problem}

\author{Daniel J.\ H.\ Chung}\thanks{Electronic mail: djchung@umich.edu}
\affiliation{Randall Physics Laboratory,
     University of Michigan, Ann Arbor, Michigan \ 48109-1120}

\author{Edward W.\ Kolb}\thanks{Electronic mail:  
rocky@rigoletto.fnal.gov}
\affiliation{NASA/Fermilab Astrophysics Center, Fermi
     National Accelerator Laboratory, Batavia, Illinois \  60510-0500,\\
     and Department of Astronomy and Astrophysics, Enrico Fermi Institute, \\
     The University of Chicago, Chicago, Illinois \ 60637-1433}

\author{Antonio Riotto}\thanks{Electronic mail: riotto@cibs.sns.it}
\affiliation{Scuola Normale Superiore, Piazza dei Cavalieri 7,
	I-56126 Pisa, Italy,\\ and INFN, Sezione di Pisa, I-56127  
	Pisa, Italy}

\date{August, 2000}

\begin{abstract}
There is no known fundamental reason to demand as a cosmological
initial condition that the bulk possess an $SO(3,1)$ isometry.  On the
contrary, one expects bulk curvature terms that violate the $SO(3,1)$
isometry at early epochs, leading to a violation of Lorentz invariance
on our brane.  Demanding that the Lorentz noninvariant terms are small
leads to a new ``flatness'' problem, not solved by the usual
formulation of inflation.  Furthermore, unlike in four dimensions, the
Lorentz violations induced from the bulk curvature cannot always be
removed as the infrared cutoff is taken arbitrarily large.  This means
that the equivalence principle in higher dimensions does not guarantee the
equivalence principle in dimensionally reduced theories.  Near-future
experiments are expected to severely constrain these
Lorentz-violating ``signatures'' of extra dimensions.

\end{abstract}


\maketitle

\section{INTRODUCTION}

Theories with extra dimensions have long played a role in attempts to
unify other forces with gravity \cite{historical}.  Traditional ideas
about hiding extra dimensions involved making them compact and small
\cite{small} (generally assumed to be of the order of the Planck
length \cite{foot1}), so that propagation of standard model matter in
the extra dimensions requires energy of the inverse of the size of the
extra dimensions.  Thus, if the extra dimensions are small enough,
they effectively decouple from the low-energy theory.

The mechanism of confining standard-model fields on
$(3+1)$-dimensional subspaces (3-branes, or just branes) \cite{brane}
of a higher-dimensional manifold leads to the possibility of scenarios
with large extra dimensions.  In models where the spacetime geometry
is of a simple factorizable form, the space of extra dimensions (the
bulk) may be compact and perhaps as large as a millimeter \cite{add}.
If the spacetime geometry has a nonfactorizable form, the extra
dimensions may be warped and noncompact as in the work of Randall and
Sundrum \cite{rs}.  The possibility of warped noncompact extra
dimensions has extended our intuition about how extra spatial
dimensions are manifest in four-dimensional effective field theories
by showing that even if gravity propagates in non-compact higher
dimensional spaces, four-dimensional observers may still empirically
deduce a four-dimensional Newton's law.

There has been a great deal of recent activity studying various
aspects of cosmology in large extra dimension scenarios.  Nonetheless,
model building is still in its infancy and general features are
still being uncovered.  As a contribution to this effort, we examine
here whether these large extra dimension scenarios possess an analog
of the ``flatness problem'' existing in four-dimensional
Friedmann--Robertson--Walker (FRW) cosmology.  We find that there is
most likely a higher dimensional flatness problem of character
significantly different from that of the FRW flatness problem.
Furthermore, unlike the flatness problem in an FRW cosmology, we will
argue this problem is not easily solvable by inflation.

We do not present our analysis in the context of any concrete
realistic model. As there is no unified theory that can address
the question of initial conditions, our conclusions necessarily must
be based on certain (plausible) assumptions.  It is impossible to know
if the fundamental theory will somehow naturally circumvent the
difficulty we discuss.  Also, we only address the issue of large (even
infinitely extending) extra dimensions.  We know that if the extra
dimensions are macroscopic, effective field theories will be valid to
describe the spacetime behavior and the calculations should be
reliable.

In the rest of the Introduction, we shall lay out the assumptions
under which our arguments apply.  However, we first discuss the
FRW flatness problem in a way that most closely parallels the flatness
problem of large extra dimensions.

One way of viewing the FRW flatness problem is as a fine-tuning
problem, associated with the fact that if the equation of state obeys
$\rho+3p>0$, {\it any} deviation from spatial flatness in the early
universe would, in a few expansion times, cause the universe either to
collapse or to expand and become negative curvature dominated.  Even
an initial condition of exact flatness is problematic since spatially
flat patches much larger than Planck size are unnatural because the
Planck scale sets the energy scale associated with quantum
gravitational fluctuations early in the universe.  That is, even if the
universe were initially globally FRW and perfectly spatially flat, any
fluctuations would have destabilized spatial flatness early on,
especially when the curvature of the spacetime was large enough for
quantum gravitational fluctuations to be large.

Note that even without introducing the issue of quantum gravitational
fluctuations there is the question of why the universe chose zero
spatial curvature over a region that is much larger than the Planck
scale to such a large degree of accuracy.  There seem to be two
options: (i) the FRW universe was nucleated by some quantum
cosmological process to have exactly zero spatial curvature, or (ii) the
FRW universe arose from a Planck-size flat patch which inflated to
become our universe.  The first of these options is possible, but
not under computational control.  The second of these options, usually
referred to as inflation, is under computational control.  The
attraction of inflation, besides computational control, is that a
physical mechanism to generate a large flat patch out of a tiny flat
patch simultaneously solves other cosmological fine-tuning problems.
It should be noted that even if inflation did not solve these other
cosmological fine-tuning problems, the solution to the flatness
problem alone would be enough reason to consider it seriously since
there is no other known physical mechanism that can solve the flatness
problem.

One may understand the FRW flatness problem as due to the fact that
there is no symmetry that prefers flat spatial geometries, and even if
the spatial geometry was spatially flat, there is no symmetry to
protect the flatness.  If one states the FRW flatness problem in those
terms, it is easy to appreciate the problem we discuss.  We will point
out that Lorentz invariance on our brane requires that the {\em
entire} spacetime possess an approximate $SO(3,1)$ isometry (not just our
brane). Since bulk curvature would lead to $SO(3,1)$ violation
observed on our brane, in the absence of some symmetry there must be
some cosmological mechanism to flatten the bulk.

The relation between bulk flatness and Lorentz invariance is easier to
appreciate if one notes that even in four-dimensional FRW models, the
reason we observe Lorentz invariance today can be connected to the
fact that our universe is flat and old.  One begins to understand this
by noting that the equivalence principle of general relativity, which
protects Lorentz invariance of the ultraviolet (UV) limit of a field
theory living on a smooth Lorentzian manifold, does not protect {\em
infrared (IR)} physics from obtaining what appears to be Lorentz
violating terms.  This statement is in some sense obvious from the
fact that field solutions generically may break the symmetry of the
underlying spacetime (a type of spontaneous symmetry breaking).  In
fact, the curvature of spacetime generated in a four-dimensional FRW
universe breaks the Lorentz isometry of the zero-energy vacuum
spacetime.  Let us see this explicitly.  If the FRW metric is written
as usual as
\begin{equation}
ds^{2}=dt^{2}-a^{2}(t)d\vec{x}^{2},
\end{equation}
we can write the action of a free scalar field in this background as
\begin{equation}
S = \int d^4\!x \ a^{3}(t)\ (\partial \phi )^{2} = \int
 d^4\!x \left[1+3H_{0}\Delta t+...\right] \left[(\partial _{0}\phi )^{2} -
(1-2H_{0}\Delta t+...) (\partial _{i}\phi )^{2} \right] , 
\end{equation}
where $H_{0}$ is the present ($t=t_{0}$) expansion rate and $\Delta
t\equiv t-t_{0}$. The different coefficients of $(\partial_0\phi)^2$ and
$(\partial_i\phi)^2$ signal Lorentz violation.  Clearly if the IR cutoff
for the field theory is taken to be $\Lambda _{IR}\gg H_{0} $, then
$H_0\Delta t \ll1$, and one can work with a Lorentz invariant field theory
\cite{foot2}. In other words, the field theory on the tangent space of the
manifold is Lorentz invariant only if the IR cutoff is taken to be larger
than $H_0$.  Of course in this four-dimensional FRW case this field theory
is really only valid on large scales (small UV energy cutoff), since the
FRW metric is only valid on large scales during most of the history of the
universe.

The only reason we ever deduced Lorentz invariance in the first place
is because the spacetime curvature associated with the energy density
in our universe is much smaller than the energy scales associated with
our physics experiments involving the standard model of particle
physics: we just can't probe energy scales smaller than $H_0$ (a similar
argument applies to planetary curvature, etc.). The
reason energy scales associated with our physics experiments are small
compared to the FRW spacetime curvature scale (the Hubble expansion
rate $H_0$) is because the universe is old, i.e., the flatness problem
\cite{foot3}.  Since inflation solves the flatness problem, we can
then say that our ability to observe Lorentz invariance today has much
to do with the fact that inflation made our universe flat (neglecting
any unaddressable metaphysical issues such as the anthropic
principle).

Extending the analogy to extra dimension scenarios is not
straightforward, however, because we do not really have a standard
time-dependent model of brane/bulk cosmology including its ``birth''
\cite{braneearly}.  What we first establish in the next section
of this paper is that departures from the $SO(3,1)$ isometry for the
{\em large scale} higher dimensional background geometry will result
in Lorentz violations for {\em any} field living in the bulk
\cite{allgrav}.  We will find that the nature of Lorentz violations
from bulk curvature is significantly different from the nature of
Lorentz violations in a four-dimensional FRW universe.  Namely, we
will see that in some cases, the Lorentz symmetry is never recovered
as the IR cutoff is taken to be arbitrarily large, unlike in the 4-D
FRW case.  This means that the four-dimensional equivalence principle
does not necessarily result from a higher dimensional equivalence
principle.  This is what we will call the ``inequivalence principle.''
Hence, if we assume that the observed four-dimensional gravitational
theory is Lorentz invariant (obeys the equivalence principle) to a
high accuracy, say to scales of order $H_0^{-1}$, an explanation is
required.

As in the 4-D FRW model, two general classes of explanations exist. One is
that for some reason the quantum gravitational dynamics is driven to
initiate an $SO(3,1)$ invariant universe, and the other is that a field
theoretical dynamical mechanism exists to drive the system to approximate
$SO(3,1)$ invariance.  The first explanation cannot be meaningfully
addressed because we do not know the fundamental unified theory. Indeed,
if such a mechanism existed, then that will most likely allow an
alternative to inflation for solving the flatness problem.  We thus take
the latter approach, and make the assumption that the universe initially
was in an $SO(3,1)$ isometry violating state, which is generic since there
is no symmetry to protect $SO(3,1)$ isometry. (For instance, $N=1$ SUSY is
compatible only with $SO(3,2)$ and $SO(3,1)$ isometries, but is
generically broken with the presence of matter energy density.)  Thus we
will require spacetime expansion, which we will call bulk inflation, to
eliminate the $SO(3,1)$ violating terms.

We find that in the most straightforward inflationary scenarios, the
warp factor does not survive bulk inflation.  Furthermore, we point
out that for the large, flat, compact extra dimension scenarios,
inflation at the effective field theory approximation cannot be used
to flatten out any significant $SO(3,1) $ violating curvature because
that would force the initial compact dimension size to be too small,
i.e., smaller than the fundamental Planck length.  This latter point
has been addressed to a certain extent in Ref.\ \cite{linde}.  Finally,
we find that $SO(3,1)$ violating curvature terms in the bulk can be
experimentally constrained.

The order of presentation will be as follows. We first discuss how
apparent Lorentz violations arise in effective field theories in a
curved spacetime in higher dimensions.  We then discuss in detail why
inflation generically would be necessary in noncompact warped extra
dimension scenarios. Finally, we discuss the possibility of finding
evidence for extra dimensions through experimental observation of
Lorentz violations and conclude.

\section{THE INEQUIVALENCE PRINCIPLE}

It is well known that in four dimensions, the IR limit of any effective
field theory is sensitive to the background spacetime curvature, which
generically induces Lorentz violations.  Due to the equivalence principle,
this Lorentz violation effect generically vanishes in the limit that the
IR cutoff is taken to be large. However, what is not as well known (at
least it is new to us) is that for any single dimensionally reduced field,
increasing the IR cutoff will not lead to the recovery of Lorentz
invariance if the $SO(3,1)$ violating curvature is due to warping in the
extra dimensions.  Hence, $SO(3,1)$ violating curvatures from extra
dimensions induce a four-dimensional theory that violates the principle of
equivalence. This is what we have referred to as the inequivalence
principle. Another way of stating the inequivalence principle is that the
equivalence principle of a higher dimensional gravitational field
theory does not necessarily guarantee that the equivalence principle
will be manifest for the dimensionally reduced effective field theory.

A perhaps noteworthy observation is that it is only an accident that
we, as four-dimensional low energy observers, discovered Lorentz
invariance as a fundamental symmetry of nature.  If the total number
spacetime dimensions is four and if we had been unlucky, spacetime
curvature would have prevented us from ever deducing Lorentz
invariance until our experiments reached an energy level above the
curvature scale.  As we will detail below, the situation in the brane
scenario could have been worse.  The inequivalence principle tells us
that if we had been unlucky to be embedded in a higher
dimensional spacetime with no $SO(3,1)$ isometry, it would have been
very difficult to deduce that gravitons obey Lorentz invariance, {\em
even at short distances} (short compared to the background curvature
scale). This would be true irrespective of the energy probed by
experiments, even if the energy were higher than the background
curvature energy scale.  This loss of Lorentz invariance in the UV
limit is a signature of extra dimensions; it cannot be reproduced with
any purely four-dimensional background curvature.
 
The loss of Lorentz invariance is connected to dimensional reduction. To
see this, consider a higher dimensional spacetime. Finding a
four-dimensional effective field theory description of a field propagating
in higher dimensions usually requires introducing an infinite number of
four-dimensional fields (a ``KK tower'') and integrating over the
coordinates of the extra dimensions.  Let us denote these four-dimensional
fields by $X_{n}$.  The procedure for obtaining a four-dimensional
effective field theory preserves the isometries, and thus the effective
field theory for the $X_{n}$ is Lorentz invariant if the underlying higher
dimensional spacetime possesses an $ SO(3,1)$ isometry with respect to the
four-dimensional coordinates of interest.

Now, what about the condition on the IR cutoff? An interesting feature of
the four-dimensional effective theory for $X_{n}$ is that none of the {\em
individual} kinetic terms for $X_{n}$ will recover Lorentz invariance even
if one takes the IR cutoff to be large.  This is because the Lorentz
invariant tangent space of the higher dimensional manifold is never four
dimensional! In other words, the IR cutoff scale for any single $X_{n}$
(for fixed $n$) is constrained to be be less than $1/L$ where $L$ is the
length scale of the extra dimensions, because $1/L$ is the UV cutoff scale
for $X_{n}$ (the field theory has been coarse-grained over length scales
of $1/L$).  One can describe the propagation in a five-dimensional
manifold up to resolution of $\Lambda _{UV}$ only when all the fields up
to mass $\Lambda _{UV}>1/L$ are taken together. However, each individual
$X_{n}$ never recovers Lorentz invariance.

Hence, we deduce two useful sufficient conditions for the existence of
Lorentz violation in the four-dimensional effective theory of a field
propagating in higher dimensions.  First, there must be no $SO(3,1)$
isometry in $D$ dimensions, and second, the coarse-graining length
scale (effective IR cutoff) must be much larger than the radius of
curvature $R$: i.e., $L \gg R$.

We now present toy models to illustrate the nature of Lorentz
violating terms induced from extra dimensions.  Consider a metric of
the form
\begin{equation}
\label{metricgeneral}
ds^{2}=A(t,U^{M})dt^{2}-B(t,U^{M})d\vec{x}^{2}
      +G_{MN}(U^M)dU^{M}dU^{N},
\end{equation}
where $U^{M}$ are coordinates of the bulk and the $U$ dependence of
$A$ and $B$ makes the global geometry nonfactorizable as in the model
of Ref.\ \cite{rs}.  Since we are interested in cosmological solutions
we have assumed that $A$ and $B$ are independent of the spatial
coordinates of the brane and only depend on $t$, the comoving time
coordinate of our universe.  Furthermore, consider the situation in
which we are interested in physics for energy scales much larger than
$\partial_t\ln A$ and $\partial_t \ln B$.  Then in the adiabatic
approximation we can neglect the time dependence of $A$ and $B$ and
set the time variable to a particular value $t_{0}$.  Under these
conditions, our generic manifold is approximately described by
\begin{equation}
ds^{2}=A(U^M)\,dt^{2}-B(U^M)\,d\vec{x}^{2}+G_{MN}(U^M)\,dU^{M}dU^{N},
\label{eq:genmetricadiabatic}
\end{equation}
where in general, 
\begin{equation}
A(U)\neq B(U) ,
\label{eq:lorentzviolate}
\end{equation}
because there is no fundamental symmetry imposing (or protecting) the
condition $A(U)=B(U)$. This implies that there is no $SO(3,1)$
isometry generated by the Lorentz group acting on the coordinates
$(t,\vec{x})$.

Although the Randall-Sundrum metric containing a single extra dimension
with coordinate $u$ ($ds^2=e^{-2b|u|}(dt^2-d\vec{x}^2) - du^2$) has an
$SO(3,1)$ isometry ($A(u)=B(u)$), the question to address is why did the
metric evolve to this form.  Brane solutions in which the bulk is not
$SO(3,1)$ isometric were recently constructed in Ref.\ \cite{bowcock},
where the metric is given by 
\begin{equation} 
ds^2=h(u)\ dt^2 - u^2
d\vec{x}^2 - h^{-1}(u)\ du^2, 
\end{equation} 
with $h(u)=-\Lambda u^2/6 - \mu/u^2$, where $\Lambda$ is a
cosmological constant in the bulk and $\mu$ is a free constant
parameter.  If $h(u)\neq u^2$, then this metric breaks the $SO(3,1)$
isometry \cite{moving}.

Now consider a simple toy model that illustrates the breaking of the
$SO(3,1)$ isometry.  Assume we live in five dimensions and the extra
spatial dimension has the topology of $S^{1}/Z_{2}$ with the radius of
$S^{1}$ equal to $L/\pi$. Suppose the background stress energy is arranged
to give the line element 
\begin{equation} 
\label{metric} 
ds^{2}=dt^{2} - e^{-2bu}d\vec{x}^{2} - du^{2}, 
\end{equation} 
where again, $u$ is the extra dimension coordinate \cite{static}. Note
that this metric is not diffeomorphic to that written by Randall and
Sundrum \cite{rs}; notably, it is not a conformally flat spacetime. It
is a static version of the cosmological spacetime presented in Ref.\
\cite{horizonevasion}.

Now suppose that a free bulk scalar field $\phi$ lives in the
background of this spacetime. We will assume that its contribution to
the vacuum energy has been fine-tuned to zero (i.e., this is not the
bulk field determining the background geometry of the spacetime), and
consider what this field looks like to a four-dimensional
observer living on our brane. The action for this bulk scalar field is
\begin{equation}
\label{5dscalar}
\Delta S_{\rm bulk}=\int d^5\!x\ \sqrt{g}\ \frac{1}{2}(\partial  
\phi )^{2},
\end{equation}
where $g_{\mu \nu}(u)$ is the $(4+1)$-dimensional metric of Eq.\
(\ref{metric}). We can consider what a four-dimensional observer will
see by expanding this scalar field in a particular orthogonal basis
\begin{equation}
\phi =\sum _{m}Y_{m}(\vec{x},t)h_{m}(u),
\end{equation}
where the $h_m$ satisfies
\begin{equation}
\frac{1}{\sqrt{g}}\ \partial _{u}[\sqrt{g}\partial  
_{u}h_{n}]=-m_{n}^{2}h_{n},
\end{equation}
and the self-adjoint derivative condition implies
\begin{equation}
\partial _{u}h_{m}|_{\rm{brane\ i}}=0 .
\end{equation}
The basis functions $h_{n}$ satisfying these conditions can be written
down explicitly:
\begin{equation}
h_{n\neq 0}=N_{n}e^{3bu/2}\left[\frac{2\pi n}{3bL}
\cos\left(\frac{\pi n}{L}u\right)-\sin\left(\frac{\pi  
n}{L}u\right)\right].
\end{equation}
If we insist on the normalization
\begin{equation}
\int du \ e^{-3bu}\ h_{n}h_{m}=\delta _{nm},
\end{equation}
we have the normalization constant
\begin{equation}
N^{-1}_{n\neq 0}=\sqrt{\frac{L}{2}}
\sqrt{1+\left(\frac{2\pi n}{3bL}\right)^{2}}.
\end{equation}

Putting this expansion into Eq.\ (\ref{5dscalar}), we find the
effective action ($\vec{\nabla }^{2}\equiv \sum_{i}
\partial_i^2$)
\begin{eqnarray}
\Delta S_{\rm bulk} & = & \int d^4\!x \ \frac{1}{2}Y_{0}
\left(-\partial _{0}^{2}+\alpha _{00}\vec{\nabla  
}^{2}-m_{0}^{2}\right)Y_{0}
\nonumber \\
&  & + \sum_{n\neq 0}\ \left[ \int d^4\!x \ Y_{n}
\left(-\partial _{0}^{2}+\alpha _{nn}\vec{\nabla  
}^{2}-m_{n}^{2}\right)Y_{n}
+ \sum _{m\neq n}Y_{m}\alpha _{mn}\vec{\nabla }^{2}Y_{n}\right] ,
\label{dimreduced}
\end{eqnarray}
where we have defined an infinite dimensional matrix
\begin{equation}
\alpha _{rn}\equiv \int du \ e^{-bu}h_{m}(u)h_{n}(u)\neq \delta_{rn}
\end{equation}
which characterizes the Lorentz noncovariant structure of the theory.
Note that for a fixed index $r$ and a given field $Y_{r}$, only the
off-diagonal components of $\alpha_{rn}$ seem to be responsible for
the Lorentz non-covariant structure, because for the diagonal
component we can always rescale the coordinates to obtain the usual
covariant form.  However, because the standard model fields confined
to the brane reveals the ``true geometry'' of the underlying
spacetime, it is not true that the field redefinition completely hides
the apparent Lorentz violation even for a given diagonal sector.  For
example, if $r=0$ we can take $x \rightarrow x\sqrt{\alpha _{00}}$ in
the first line of Eq.\ (\ref{dimreduced}) to obtain
\begin{equation}
\label{zeromodeeq}
S_0^{\rm{scaled}}=\int d^4\!x \ \alpha _{00}^{3/2}\ \frac{1}{2} \
Y_{0}\left(-\partial _{0}^{2}+\vec{\nabla }^{2}-m_{0}^{2}\right)Y_0,
\end{equation}
apparently recovering Lorentz invariance for what we will call the
$00$ sector.

It is manifest that two points that are a distance $d$ apart (as measured
by the standard model physics of fields confined to our brane) are seen to
be only a distance $d/\sqrt{\alpha _{00}}$ apart from the point of view of
the effective four-dimensional field, which actually lives in higher
dimensions.  Of course there is no global coordinate transformation that
one can make to have all diagonal $nn$ sectors Lorentz covariant
simultaneously. Hence, at least with the off-diagonal terms neglected,
each field $Y_{n}$ lives in a different apparent geometry; i.e., the
effective distance that each field sees through its propagator is
different even though the ``true'' distance in spacetime as measured by
the standard model fields confined to the brane is the same.  The distance
$d/\sqrt{\alpha _{00}}$ is what an observer would deduce from an
``inverse-square law'' analysis, and hence we will refer to it as the
``inverse-square distance.'' The ratio of the inverse-square distance to
the distance measured by fields confined to the brane is in this case just
$1/\sqrt{\alpha_{00}}$.  The fact that the ratio is not unity is nothing
more than a consequence of the fact that the lightcone in the
extra-dimensional spacetime is different from the lightcone of a field
confined to the brane, as was discussed in Ref.\ \cite{horizonevasion}. In
other words, causal signals can take a shortcut through the extra
dimensions to get to a point on the brane that is farther than where a
causal signal confined to the brane can go for a fixed time. From a
$(3+1)$-dimensional point of view, the higher dimensional signals seem
acausal.

It is important to note that this noncovariant structure is
independent of the basis chosen, and there is no coordinate
redefinition nor field redefinition that will truly restore the Lorentz
symmetry. Even more importantly, since the Lorentz violation structure
is governed by the quantity $bL$ (which is independent of
$\Lambda_{IR}$, the IR cutoff of the $3+1$ dimensional momenta),
increasing $\Lambda_{IR}$ does not lead to the recovery of Lorentz
invariance for any one field $Y_n$.  Hence, the inequivalence
principle is manifest.  As argued before, the fact that $\alpha
_{mn}\neq \delta _{mn}$ is a result of the fact that the underlying
higher dimensional spacetime does not possess an $SO(3,1)$
isometry. Mathematically, this merely amounts to the fact that the
partial differential equations governing the modes are not separable
in the chosen coordinate directions.

Let us now examine the magnitude of these effects. The magnitude of the
Lorentz-violating effects is characterized by the coefficients $\alpha
_{ij}$. First note that in this model the Lorentz violating effects
associated with the zero mode are not very large, because $1\leq \alpha
_{00}\leq 3$. In particular, the distances are only scaled by
$d/\sqrt{\alpha_{00}}$. However, since the scalar field propagators behave
approximately as the graviton propagator for Newton's law, one can see
that the ``inverse-square'' distance vs.\ luminosity distance comparisons
can discriminate such scalings \cite{spintwo}.  We leave a more careful
analysis of the observables to a future study.  However, in the last
section, we will explicitly show one possible experimental observable
which is within the reach of upcoming gravitational experiments.

The effects for the higher mass mode can be extremely large, even if
$L^{-1}$ is much larger than electroweak energy scale. For instance, if
$e^{bL} \gg 1$, we have 
\begin{equation} 
\alpha _{11}\sim e^{2bL}\frac{29\pi ^{2}}{18b^{3}L^{3}} , 
\end{equation} 
which will be much larger than unity. Moreover, it is not always
possible to treat the mixing of the zero-mode mass eigenstates with
massive KK mass eigenstates as a perturbation because the mixing with
massive modes can be equally large if $e^{bL}$ is large, as can be
seen by
\begin{equation} 
\alpha _{01}\sim e^{bL/2}\frac{16\pi }{(bL)^{2/3}}\sqrt{\frac{2}{3}} .
\end{equation}
In general, the zero mode truncation of the effective field theory in
the $bL \gg 1$ limit is not valid because $\alpha _{mn}$ is much
greater than unity if $bL$ is much greater than unity, and the field
theory must be considered from a higher dimensional point of view.
This may be true even though the nonzero modes (the zero mode is
massless) can be quite massive since $L^{-1}$ may need to be large
enough to hide the higher dimensional behavior of gravity. Explicitly,
the mass spectrum of the nonzero modes is given by
\begin{equation}
m_{n\neq0}=\frac{1}{2}\sqrt{9b^{2}+\left(\frac{2\pi n}{L}\right)^{2}} ,
\end{equation}
which would naively justify decoupling if $b$ or $L^{-1}$ were
sufficiently large. However, here in general, the mass eigenstates
will not be momentum eigenstates, and there does not seem to be
decoupling. What is clear, however, is that if we insist on a
four-dimensional point of view, we have a theory in which the field
labeled by different masses see a different effective geometry, i.e.,
the inverse-square distances corresponding to the same spacetime
geometry distance are different for different four-dimensional
effective fields.

Although the exact nature of the Lorentz violating effects
characterized by $\alpha _{mn}$ is model dependent, its magnitude
can be read off from the metric of the form given in Eq.\
(\ref{metricgeneral}).  It is easy to show that in general whether
$\alpha _{mn}$ is greater than or less than unity is roughly governed
by the ratio
\begin{equation}
\alpha _{mn}\sim \frac{\langle B\rangle }{\langle A\rangle} ,
\end{equation}
where $\langle \cdots \rangle$ denotes an average over the extra
dimensions. Hence, for spacetimes with $\langle B\rangle/\langle
A\rangle>1$ we have an ``acausal'' effective theory while for $\langle
B\rangle/\langle A\rangle <1$ , we have merely the momentum
nonconserving Lorentz violating effects with respect to the off-diagonal
terms of $\alpha _{mn}$.

Note that the existence of Lorentz violation generalizes to higher
spin fields. Consider the graviton field $h_{\mu\nu}$, which is
defined to be the zero mode of the metric perturbation
\begin{equation}
ds^2 =A\left[(1+h_{00}) dt^{2}
- \left(\frac{B}{A}\delta_{ij}-h_{ij}\right)dx^idx^j\right]
+ G_{MN}dU^{M}dU^{N} ,
\end{equation}
where $h_{\mu\nu}$ only depends on $(3+1)$-dimensional
coordinates. The graviton kinetic term will contain a term
\begin{equation}
S  \ni  \int d^n\!x \ \sqrt{AB^3}\ \sqrt{G}\  \frac{1}{4}
\left[A^{-1} (\partial_0h^{\mu \nu} \partial_0  h_{\mu \nu})
- B^{-1}(\vec{\nabla}h^{\mu \nu} \cdot \vec{\nabla}  h_{\mu \nu})  
\right] ,
\label{eq:gravitonkinetic}
\end{equation}
which is again Lorentz violating when integrated over the extra
dimension coordinates $U$.  Again, the metric implied by measurements
of the brane-confined fields will be different from the constant
metric obtained after integrating over $U$ in Eq.\
(\ref{eq:gravitonkinetic}).

\section{A new flatness problem}

Since we have little control over the effective field theory in the
context in which the extra dimensions are compactified to be Planck
size, we will not discuss that scenario here. However, in the case in
which the extra dimensions are large and flat \cite{add} or in the case
in which the extra dimensions are warped and noncompact \cite{rs}, we
can ask in the context of an effective field theory whether there may
be a flatness problem as outlined in the Introduction.  As we shall
argue, in these large extra dimension scenarios there are additional
flatness problem complications that did not arise in the
four-dimensional FRW cosmology. In what follows, we shall first
identify the flatness problem in the warped extra dimensions scenario
and then discuss the case of the large extra dimension scenario.

As discussed more fully in the previous section, a higher dimensional
spacetime described by Eqs.\ (\ref{eq:genmetricadiabatic}) and
(\ref{eq:lorentzviolate}) implies that any dimensionally reduced
effective theory, including gravity, will generically violate Lorentz
invariance. In the view of treating gravity as a theory of vielbeins
in Minkowski spacetime, this means that the VEV of the
vielbeins spontaneously break Lorentz invariance. Hence, in warped
extra dimensions scenarios, it is crucial to explain why there is an
$SO(3,1)$ isometry in the extra dimensions to an accuracy that allows
four-dimensional gravity (or any other dimensionally reduced field) to
be Lorentz invariant \cite{vacuum}.  Hence, in analogy to the FRW
flatness problem, if we assume that the bulk fields dimensionally
reduced to four dimensions are observed to obey Lorentz invariance (as
the IR cutoff is taken to be arbitrarily large), we have an
extra-dimensional flatness problem.  This analogy is summarized in Table 1.

\begin{table}
\caption[t1]{\label{sigmas} The analogy between the FRW flatness problem 
and the extra-dimension flatness problem.}
\begin{ruledtabular}
\begin{tabular}{cc}
4D FRW & Extra dimensions \\
\hline  \hline 
no reason for initial spatial flatness & 
                no reason for initial $ SO(3,1)$ isometry \\
spatial curvature $\rightarrow$ dynamical instability & ? \\
observation: old age of the universe & 
                observation: approximate $SO(3,1)$ isometry \\
\end{tabular}
\end{ruledtabular}
\end{table}

If a typical initial condition of early cosmology contains $SO(3,1)$
isometry violating curvature, the extra dimensional flatness problem
is real and inflation may be required to eliminate it.  Note that here
we are making a crucial assumption that there is no fundamental reason
(such as fundamental symmetry arguments or dynamical arguments) to
choose an $SO(3,1)$ isometric manifold as the initial condition.
Indeed, if there were such a mechanism, then one may be able to modify
it and utilize it to replace inflation altogether.

Given that we use inflation to solve the FRW flatness problem,
let's see what normal inflation would do to solve the bulk flatness
problem. Let's extend the toy model of the previous section by
allowing a brane scale factor $a(t)$:
\begin{equation}
ds^{2}=dt^{2} - a^{2}(t)\ e^{-2bu}\ d\vec{x}^{2} - du^{2},
\end{equation}
where again, $u$ is the extra dimension coordinate and our flat
brane is located at $u=0$.  It is easy to see that no matter how much
we inflate our brane by arranging $a(t)$ to grow exponentially, we
will not recover the $SO(3,1)$ isometry.

One possible resolution to this problem is to inflate the $u$
dimension. The difficulty lies in inflating the extra dimensions to
smooth out the $SO(3,1)$ invariance violating curvature while
preserving the large warping.  To see this, introduce a bulk scale
factor $c(t)$ for the extra dimension.  The toy metric is then
\begin{equation}
ds^{2}=dt^{2}-a^{2}(t)\ e^{-2bu}\ d\vec{x}^{2}-c^{2}(t)\ du^{2} .
\label{eq:beforeinflation}
\end{equation}
Inflating $c(t)$ (``bulk inflation'') will render the curvature set by
$b$ harmless.  One can see this by making the coordinate
transformation $\tilde{u}=cu$, in which case the metric becomes
\begin{equation}
ds^{2} =  
\left[1-\left(\frac{\dot{c}}{c}\right)^{2}\tilde{u}^{2}\right]dt^{2}
- a^{2}(t) \ e^{-2b\tilde{u}/c}\ d\vec{x}^{2} -d\tilde{u}^{2}
+ 2 \left(\frac{\dot{c}}{c}\right)\tilde{u}\ d\tilde{u} \, dt .
\label{eq:inflated}
\end{equation}
Now imagine $c$ inflates by large amount, after which
$\dot{c}/c\rightarrow 0$ (or at least $\dot{c}/c<\dot{a}/{a}$).  Then
the factor $b/c$, which sets the spatial curvature scale in the bulk,
can be made arbitrarily small while still recovering $SO(3,1)$
isometry.  Of course the price one pays here is that inflation of the
bulk has inflated away the warp factor!

An obvious loophole in the argument thus far is the possibility of a
hierarchy between the Lorentz violating curvature and the warp factor
curvature.  Then one may be able to inflate away the Lorentz violating
curvature without erasing the warp factor.  For example, if one
complicates the toy metric a bit further and takes it to be
\begin{equation}
ds^{2}=e^{-2ku}\left[dt^{2}-a^{2}(t) \ e^{-2bu} \ d\vec{x}^{2}\right] 
- c^{2}(t) \ du^{2}
\end{equation}
with $k\gg b$, then bulk inflation may dilute away the curvature due
to $b$ while maintaining the warp factor from $k$.  The challenge then
is to come up with a physical scenario with a large hierarchy between
$k$ and $b$.  Of course, there may be other solutions that involve the
bulk inflating and then shrinking in such a way that the $SO(3,1)$
violating curvature is removed, but the main point still stands:
inflationary model building has a new problem to solve.

Finally, consider the large compact extra dimension scenarios
\cite{add} with $r$ extra spatial dimensions.  Imagine that an
$SO(3,1)$ violating metric of the form Eq.\ (\ref{eq:beforeinflation})
has been inflated as in Eq.\ (\ref{eq:inflated}) to end with an
acceptably flat, compact extra dimension.  Let us parameterize such a
metric as
\begin{equation}
ds^{2} = dt^{2} - a^{2}(t)e^{-2bu}d\vec{x}^{2}
- c^{2}(t)\delta_{MN}dU^{M}dU^{N}
\end{equation}
where $U^{M}$ correspond to the compact extra dimensions where $M$
runs through 4 to $3+r$. If the extra dimensions described by the
coordinates $U$ are compact, then inflation through $c(t)$ poses the
danger of the compact dimensions being initially too small to be
described by an effective field theory. More specifically, let us
define the total expansion of the extra dimensions between some initial
time $t_{i}$ and today ($t_{0}$) as
\begin{equation}
E = \frac{c(t_{0})}{c(t_{i})} = \frac{b}{H_{0}} ,
\end{equation}
where the second equality is required by the requirement that the
curvature not violate $SO(3,1)$ invariance out to scale of
$H_{0}^{-1}$. Suppose the size of the extra dimensions is initially
$l_{i}$. After inflation, the size of the extra dimensions is
$l_{0}=l_{i}E$.  Then since the four-dimensional Planck scale requires
$l_{0}^{r}M^{2+r}\approx M_{Pl}^{2}$ where $M$ is the $r$-dimensional
Planck scale, we find
\begin{equation}
\left(\frac{M^2b}{M_{Pl}^2H_0}\right)^{1/r} M=\frac{1}{l_{i}}.
\end{equation}
If we require an effective field theory description to be valid
by imposing $ l_i^{-1}\leq M $, we find
\begin{equation}
b\leq H_{0}\left(\frac{M_{Pl}}{M}\right)^{2}.
\end{equation}
This means that if we require that the fundamental Planck scale $M$ to
be $M\geq 1$ TeV, we find that the curvature in the bulk before bulk
inflation can be only very tiny, $b\leq 10^{-1}$eV.  Hence, we find
that for large {\em compact} extra dimensions, only very tiny
curvatures can be smoothed out by inflation.  This suggests that a
non-quantum-gravitational theory of inflation cannot smooth away the
bulk curvature, and most likely some fundamental symmetry must
play a role in initially setting the $SO(3,1)$ violating curvature to
zero.  A related discussion can be found in Ref.~\cite{linde}.

Let us reiterate the main point of this section which is the main
result of this paper. For {\em noncompact} warped
extra dimension scenarios, one must invoke a mechanism such as
inflation to smooth out the $SO(3,1)$ isometry violating curvature to
obtain a Lorentz invariant effective field theory.  However, the
difficulty with this resolution is that in such an inflationary
scenario, the bulk warp factor that one needs to localize gravity may
be inflated away altogether. Hence, the new flatness problem is to
inflate away the $ SO(3,1) $ violating curvature selectively while
preserving a large warp factor.  In the case of large compact extra
dimensions, we find that inflation at the level of an effective field
theory generically cannot make the bulk flat.

\section{Conclusion}

We have shown that curvature in the bulk leads to an apparent breaking
of Lorentz symmetry with respect to an observer living on the brane
observing a field propagating in higher dimensions. Unlike the
$SO(3,1)$ violation in the four-dimensional world due to curvature,
Lorentz violation in theories with fields propagating in higher
dimensions persists as long as the spacetime does not possess an
$SO(3,1)$ isometry, no matter how large of an energy the
four-dimensional effective field theory is probed.  This is in
contrast with what is dictated by the equivalence principle in
four dimensions.  We called this apparent UV-limit-persisting
violation of four-dimensional equivalence principle for dimensionally
reduced theories the ``inequivalence principle.''  Note that Lorentz
violation for the dimensionally reduced theory is true even when the
brane has an $SO(3,1)$ isometry.  Furthermore, the mismatch between
the brane isometry group and the full spacetime isometry group results
in an ambiguous geometrical picture from a four-dimensional empirical
point of view regardless of the symmetry group.

This implies a new flatness problem for cosmological scenarios having
large (possibly noncompact) extra dimensions. For warped noncompact
extra dimensions scenarios, the problem is to come up with a mechanism
to flatten the $SO(3,1)$ violating wrinkles while preserving the large
warp factor necessary for graviton trapping. Inflation generically
helps to smooth away the wrinkle, but it also eliminates the warp
factor. For flat compact large extra dimensions scenarios, we
demonstrated that inflation at the effective field theory level is not
sufficient to smooth away any significant curvature in the bulk.

It is tempting to speculate that the first signatures for large extra
dimensions may come from deducing the existence of Lorentz violations
in the early universe. This would be possible only if the observable
anomalies in the early universe arising from Lorentz violations are
sufficiently distinct from the anomalies arising from other
effects. Indeed, since in most popular scenarios only gravity and
other extremely weakly interacting (with the SM fields) fields
propagate in the bulk, it may be difficult to observe the Lorentz
violations with respect to the zero modes of these fields unless the
violation is extreme. 

One Lorentz violating observable characteristic of the inequivalence
principle is the wavelength independent deviation of signal
propagation speed.  For example, for the free scalar field of
Eq.\ (\ref{5dscalar}) propagating in a five-dimensional background of
Eq.\ (\ref{metric}), one can solve the wave equation perturbatively,
perturbing with the parameter $b L$ where $b$ characterized the
Lorentz violating curvature and $L$ is the characteristic size of the
extra dimension.  One can write one of the modes as
\begin{equation}
\phi \sim \frac{f(u)}{a}\ e^{-iE_k t} \ e^{i\vec{k}\cdot{\vec{x}}} ,
\end{equation}
where
\begin{equation}
f=c_0\left[1+ b\vec{k}^2 \left(\frac{u^3}{3}-\frac{L u^2}{2}\right)\right],
\end{equation}
and
\begin{equation}
E_k=\left(1+\frac{1}{2} bL\right)|\vec{k}|.
\end{equation}
This implies that the group velocity is 
\begin{equation}
\frac{\partial E_k}{\partial |\vec{k}|} =1+\frac{bL}{2} ,
\end{equation}
in agreement with Eq.\ (\ref{dimreduced}) since there we have 
\begin{equation}
\frac{\partial E_k}{\partial |\vec{k}|} = \sqrt{\alpha_{00}}.
\end{equation}
As we have argued before, the propagation speed of gravitational waves
will be similar.  Hence, we conclude that one may be able to constrain
the bulk-curvature violating $SO(3,1)$ isometry by comparing the
gravitational wave arrival time and the light arrival time.  

Imagine measuring the time correlation of the arrival of the
gravitational wave and light pulses from a gamma-ray burst.  Taking
the gamma-ray burst to be at a distance $D=1000$ Mpc $\sim 10^{17}$ s,
and assuming a resolution for the arrival times of the pulses of
$\Delta t = 1$ s, we would be able to constrain the Lorentz violating
curvature to be smaller than
\begin{equation}
\frac{bL}{2} < \frac{\Delta t}{D} = 10^{-17}
\end{equation}
if no time lag is detected.

In conclusion, any realistic cosmological model with extra
dimensions must account for a mechanism to generate approximate Lorentz
symmetry for fields in the bulk, if the dimensionally reduced bulk field
such as the graviton can be shown to be approximately Lorentz invariant.
If we are lucky, perhaps nature will give us a clue regarding the
extra dimensions through tests of Lorentz violations in the graviton
sector and any other sector that may live in extra dimensions.

\begin{acknowledgements}
DJHC thanks L. Everett, G. L. Kane, J. Wang, J. Liu, J. Lu,
and R. Garisto for interesting conversations.  DJHC was supported
by the Department of Energy.  EWK was supported by the Department of
Energy and NASA under Grant NAG5-7092.
\end{acknowledgements}



\begin{thebibliography}{99}
\bibitem{foot1} 
Here the Planck length is defined to be $G_N^{1/2}$.
\bibitem{foot2}
One way to impose an IR cutoff is to put the theory in a box with
periodic boundary conditions.
\bibitem{foot3}
If the universe had even a tiny positive large-scale spatial curvature
at some early time, it would have collapsed without expanding
sufficiently.  Without sufficient expansion, the energy density would
never have sufficiently diluted to allow the curvature of spacetime to
be small compared to the electroweak scale.
\bibitem{allgrav}
We assume all fields interact with gravity.
\bibitem{moving}
Here, the brane is moving in the static bulk, but the movement can be
treated adiabatically.
\bibitem{spintwo}
Although the details resulting from the spin-two nature and the coupling
to sources will change the details between the graviton propagator and
the scalar field propagator, the qualitative behavior of the field
propagation should be the same.
\bibitem{vacuum}
Of course, it is not empirically clear
whether graviton dynamics about our ``vacuum'' obey Lorentz
invariance.
\end{thebibliography}
\end{document}